\documentclass[namedreferences]{solarphysics}

\usepackage[hyperref,optionalrh,showbiblabels]{spr-sola-addons} % For Solar Physics 
\usepackage{graphicx}        % For eps figures, newer & more powerfull
\usepackage{color}           % For color text: \color command
\usepackage{breakurl}        % For breaking URLs easily trough lines
\newcommand{\aap}{    {\it Astron. Astrophys.}}
\newcommand{\aaps}{   {\it Astron. Astrophys. Suppl.}}

\newcommand{\aj}{     {\it Astron. J.}} 
\newcommand{\apj}{    {\it Astrophys. J.}}

\newcommand{\pasj}{   {\it Pub. Astron. Soc. Japan}}

\newcommand{\solphys}{{\it Solar Phys.}}
 
\newcommand{\ssr}{    {\it Space Sci. Rev.}} 
\chardef\us=`\_

\begin{document}

\begin{article}
\begin{opening}

\title{The Na\,{\sc i} and Sr\,{\sc ii} Resonance Lines in Solar Prominences\\} 

\author[addressref={aff1},corref,email={stell@iap.fr}]{\inits{G.S.}\fnm{G. Stellmacher}}%\sep
\author[addressref=aff2,email={ewiehr@astro.physik.uni-goettingen.de}]{\inits{E.W.}\fnm{E. Wiehr}}%\sep
%\author{\inits{}\fnm{}~\lnm{}\orcid{}}
%\author{P.~\surname{Author-a}$^{1}$\sep
%        E.~\surname{Author-b}$^{1}$\sep
%        M.~\surname{Author-c}$^{2}$      
%       }

%   \institute{$^{1}$ First affiliation
%                     email: \url{e.mail-a} email: \url{e.mail-b}\\ 
%              $^{2}$ Second affiliation
%                     email: \url{e.mail-c} \\
%             }
\address[id=aff1]{Institute d'Astrophysique (IAP), 98 bis Blvd. d'Arago, 
             75014 Paris, France}
\address[id=aff2]{Institut f. Astrophysik,
             Fr.-Hund-Platz 1, 37077 G\"ottingen, Germany}

\runningauthor{G. Stellmacher and E. Wiehr}
\runningtitle{The Na\,{\sc i} and Sr\,{\sc ii} Resonance Lines in Solar Prominences}

\begin{abstract}
We estimate the electron density, $n_e$, and its spatial variation in 
quiescent prominences from the observed emission ratio of the resonance 
lines Na\,{\sc i}\,5890\,\AA{}\,(D$_{2}$) and Sr\,{\sc ii}\,4078\,\AA{}. 
For a bright prominence ($\tau_{\alpha}\approx25$) we obtain a mean
$n_e\approx2\times10^{10}$\,cm$^{-3}$; for a faint one ($\tau_{\alpha}\approx4$) 
$n_e\approx4\times10^{10}$\,cm$^{-3}$ on two consecutive days with moderate 
internal fluctuation and no systematic variation with height above the solar 
limb. The thermal and non-thermal contributions to the line broadening, 
$T_{\rm kin}$ and $V_{\rm nth}$, required to deduce $n_e$ from the emission ratio 
Na\,{\sc i}/Sr\,{\sc ii} cannot be unambiguously determined from observed 
widths of lines from atoms of different mass. The reduced widths, 
$\Delta\lambda_{\rm D}/\lambda_0$, of Sr\,{\sc ii}\,4078\,\AA{} show an 
excess over those from Na\,D$_{2}$ and H$\delta\,4101$\,\AA{}, assuming 
the same $T_{\rm kin}$ and $V_{\rm nth}$. 
We attribute this excess broadening to higher non-thermal broadening 
induced by interaction of ions with the prominence magnetic field. 
This is suggested by the finding of higher macro-shifts of  
Sr\,{\sc ii}\,4078\,\AA{} as compared to those from  Na\,D$_2$.

\end{abstract}
\end{opening}
%-------------------------------------------------

\section{Introduction}
     \label{S-Introduction} 
 
The knowledge of fundamental plasma parameters in solar prominences like 
electron density, $n_e$, and kinetic temperature, $T_{\rm kin}$, is still rather 
unsatisfactory ({\it cf.,} Labrosse {\it et al.}, 2010). The large range of $n_e$ 
values may be illustrated by the analysis of Bommier et al. (1994), who 
interpret Hanle-effect observations of H$\alpha$ and HeD$_3$ and obtain 
$0.25\times10^{10}\le n_e\le 6.3\times10^{10}$\,cm$^{-3}$ with a mean of 
$n_e= 2.1\times10^{10}$\,cm$^{-3}$. Koutchmy {\it et al.} (1983) derived from 
eclipse spectra of H$\beta$ and the Balmer continuum 
$n_e < 2.5\times10^{10}$\,cm$^{-3}$. 

Landman (1983\,a) calculated for a homogeneous atmosphere the dependence 
of the Na\,D$_2$ and Sr\,{\sc ii}\,4078\,\AA{} emissions on $n_e$, and
compared (in Landman, 1983\,b) the ratio of their integrated line emission 
with observations by Shih-Huei (1961) and by Yakovkin and Zel'dina (1964), 
deducing a mean electron density $n_e\approx6\times10^{10}$\,cm$^{-3}$. 
Considering more energy levels of Na\,{\sc i} and Sr\,{\sc ii}, Landman 
(1986) revised his 1983 results for $n_e$ by a factor of 0.5, leading 
to a mean electron density $n_e\approx3\times 10^{10}$\,cm$^{-3}$. 

Although Landman (1983\,a) found only a slight dependence on the 
line-broad\-ening parameters, these cannot be fully neglected. Thermal and 
non-thermal contribution to line-broadening, $T_{\rm kin}$ and $V_{\rm nth}$ 
cannot unambiguously be obtained from Doppler widths: Balmer, helium and 
metallic lines do not yield a common pair [$T_{\rm kin}$ , $V_{\rm nth}$]. 
Wiehr {\it et al.} (2013) found that the reduced Doppler widths 
$\Delta\lambda_{\rm D}/\lambda_0$ of lines from ions are generally too 
broad as compared to those from neutrals.

Such a selective excess broadening cannot occur in a homogeneous prominence 
plasma, where neutrals and ions behave as in a one-component fluid 
(Terradas {\it et al.}, 2015). On the other hand, Low {\it et al.} (2012) 
showed that condensations of cool gas must occur in prominences ``as a 
consequence of the non-linear coupling of force balance and energy transport''. 
This ``catastrophic radiative cooling'' causes gas clumps to become largely 
neutral, and their frozen-in state breaks down. These clumps then sink and 
progressively ``warm up by resistive heating'' until, at sufficient 
re-ionization, the clumps are eventually again frozen-in and stop sinking. 

If some fraction of this alternating `stop-and-go' falling motion shows up 
as non-thermal velocity (due to turbulence and/or shocks; {\it cf.} Hillier 
{\it et al.}, 2012), an excess broadening should occur for lines of the 
(sinking) neutrals. Yet, the observed broader lines from ions (Stellmacher 
and Wiehr, 2015), supposed to be frozen-in, rather indicate the contrary. 
We examine this scenario comparing the width of Sr\,{\sc ii}\,4078\,\AA{} 
and Na\,D$_2$, which may probe prominence regions with different excitation 
conditions. 

\section{Observations}                                 
    \label{S-obs}

With the Vacuum Tower Telescope, VTT, on Tenerife we observe four quiescent 
prominences in 2015. Two brighter ones are observed on August 31 (B) and 
September 1 (C), a medium bright one on August 30 (A) and a fainter one 
on two successive days on September 2 and 3 (D). Characteristic parameters 
for these prominences are given in Table\,1.   

The neighboring lines Sr\,{\sc ii}\,4078\,\AA{} and H$\delta$ are 
simultaneously recorded in the 55th order and Na\,D$_2$ is recorded in the 
38th order about 1\,min later after a change of the spectrograph setting, 
since the pre-selector of the VTT does not allow a simultaneous observation 
of both lines. The neighboring He\,3888\,\AA{} and H$_8$\,3889\,\AA{} lines 
are recorded in the 58th order a further minute later. The CCD exposure is 
50\,ms and the width of the spectrograph-slit of 0.55\,arcsec corresponds to 
400\,km on the Sun. We estimate the influence of seeing during the 50\,ms 
exposures to amount to 1.5\,arcsec.

%+++++++++++++++++++++++++++++++
%% Table(1)
\begin{table}[h]
\caption{Parameters of the observed prominences: designation, date, 
limb position, maximum H$\delta$ brightness [erg/(s cm$^2$ ster)], and the 
corresponding optical depths taken from the tables by Gouttebroze, Heinzel, 
Vial (1993)}. 
\begin{tabular}{cccccc} % l: left, c: center, r: right
\hline 
\hspace{-4mm} prominence&date&position& E$^{\rm max}(\delta$)&$\tau^{\rm max}(\delta$)&$\tau^{\rm max}(\alpha$)\\ 
\hline  
\hspace{-4mm} A&Aug. 30& E/10S &$13\cdot10^3$ &0.15 & 7 \\
\hspace{-4mm} B&Aug. 31& E/15S &$25\cdot10^3$ &0.29 & 13 \\
\hspace{-4mm} C&Sept. 1& W/35N &$46\cdot10^3$ &0.5  & 25 \\ 
\hspace{-4mm} D&Sept. 2& W/15S &$8\cdot10^3$  &0.09 & 4 \\
\hspace{-4mm} D&Sept. 3& W/15S &$8\cdot10^3$  &0.09 & 4 \\
\hline  
\end{tabular}
\end{table}
%+++++++++++++++++++++++++++++++

Spectra of the (emission free) prominence neighborhood are taken immediately 
before and after each exposure for a determination of the stray-light aureole
(mainly Rayleigh scattering on the mirror surfaces, {\it cf.}, Stellmacher and 
Wiehr, 1970). The change of the spectrograph setting between the two exposures 
(Na\,D, respectively, Sr\,{\sc ii} with H$\delta$) does not preserve the
wavelengths precisely on the CCDs. As a consequence, the aureole exposures 
have to be repeated for each set of emissions. This, in turn, requires that 
the telescope be pointed alternately to the prominence and to its neighborhood. 
Since the VTT occasionally does not return precisely to the same prominence 
position, we only use those Na\,D$_2$ spectra, which visually correlate with 
the corresponding Sr\,{\sc ii} and H$\delta$ spectra.
 
The integrated emissions, $E_{\rm tot}$, are calibrated in absolute units 
[erg/(s cm$^2$ ster] with spectra of the disk center, scaled with values by 
Labs and Neckels (1970); Ramelli {\it et. al.} (2012) described the reduction 
procedure in detail. We determine line profiles in spatial emission 
maxima averaged over 3 pixels of the 2-pixel binned CCDs, corresponding 
to $\approx$1\,arcsec. The estimated seeing of $\approx1.5$\,arcsec 
corresponds to a resolution area of 1000\,km$\times$1000\,km on the sun. 
Different refraction of the yellow Na\,D and the violet Sr\,{\sc ii} 
light is well below 0.5\,arsec at 2400\,m above sea level and a solar 
zenith distance of $<35^{\circ}$.  

The obtained emission line profiles are fitted by Gaussians. 
The main uncertainty arises from the accuracy of the zero level, which is 
affected by residual absorption lines from the stray-light aureole. We 
determine the zero level as broad-band median intensity outside the respective 
emission line. In order to avoid an influence of line asymmetries we 
successively fit the upper 10\%, 40\% and 70\% of each emission profile. For 
the determination of the wavelength position on the CCD, we use only the upper 
10\%, thus avoiding influences from blending by emission satellites. For a 
determination of the line widths, $\Delta\lambda_e$, and of the integrated 
line intensity, $E_{\rm tot}=\int I_{\lambda} d\lambda$, we use only symmetric 
profiles for which the three Gaussians (of 10\%, 40\%, 70\%) do not 
significantly differ. In this way we estimate the shift accuracy to be 
$\le0.5$\,pixel, corresponding to $\Delta V_{\rm macro} \le 0.01$\,km/s or 
$\Delta\lambda_{\rm e}\le2$\,m\AA{}, and $\Delta$$E_{\rm tot}/E_{\rm tot}\le0.05$.

\section{Results} 
  \label{S-results}

For several exposures the visual aspects of the Na\,D$_2$ and the 
Sr\,{\sc ii}\,4078\,\AA{} emissions are so conspicuously similar that 
it is impossible to distinguish them at first glance. This is shown in 
Figure\,1 for a set of spectra of prominence D. We reasonably assume that this 
similarity is real, and that an occasionally occurring poorer coincidence of 
both lines is due to imperfect return of the VTT from the intermediate aureole 
pointing to the precedent location in the prominence, and possibly also to 
different influence of seeing during the two exposures. 

%________________________________________________________________
%  Fig.1  [Sept03Komb]
   \begin{figure} [h]
   \hspace{-1mm}
   \includegraphics[width=0.92\textwidth]{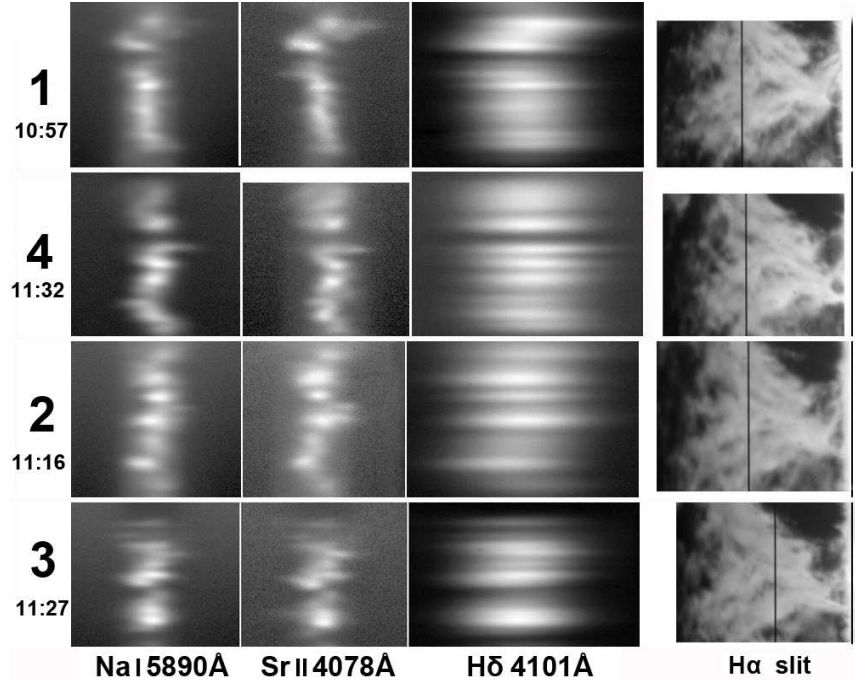}
   \caption{Spectra of NaD$_2$, Sr\,{\sc ii} and H$\delta$ for 
   prominence D at slit-positions\,1, 4, 2 and 3, decreasing with height 
   above the solar limb (black vertical lines at right border of H$\alpha$ 
   slit yaw images). Each spectrum covers 0.6\,\AA{}$\times~70$\,arcsec.}
%\label{Fig1}
    \end{figure}
%________________________________________________________________
%

\subsection{Total Emission and Electron Density}
  \label{S-Etot}

The ratio of the total line emissions, $E_{\rm tot}=\int I_{\lambda}$d$\lambda$,
of Na\,D$_2$ and Sr\,{\sc ii}\,4078\,\AA{} allows to estimate the electron 
density $n_e$. Landman (1983\,a) found that under prominence conditions the 
Sr atom is mainly ionized. He gave for the Sr\,{\sc ii} resonance line a 
Boltzmann distribution with $n$(Sr\,{\sc ii}, 5$^2P) \approx n$(Sr\,{\sc ii}) 
$\approx n$(Sr$^{tot})$, largely independent of temperature. The Na\,{\sc i} 
resonance line can only be emitted after recombination and is thus sensitive 
to $n_e$. Landman (1983) found a Saha-Boltzmann distribution with 
$n$(Na\,{\sc i},\,$3^2P)\propto n_e~n$\,(Na\,{\sc ii}) $\approx n_e~n$(Na$^{tot}$). 

%________________________________________________________________
%  Fig.2  [E_Delta./.E_Sr]
   \begin{figure}
   \hspace{-3mm}
   \includegraphics[width=0.98\textwidth]{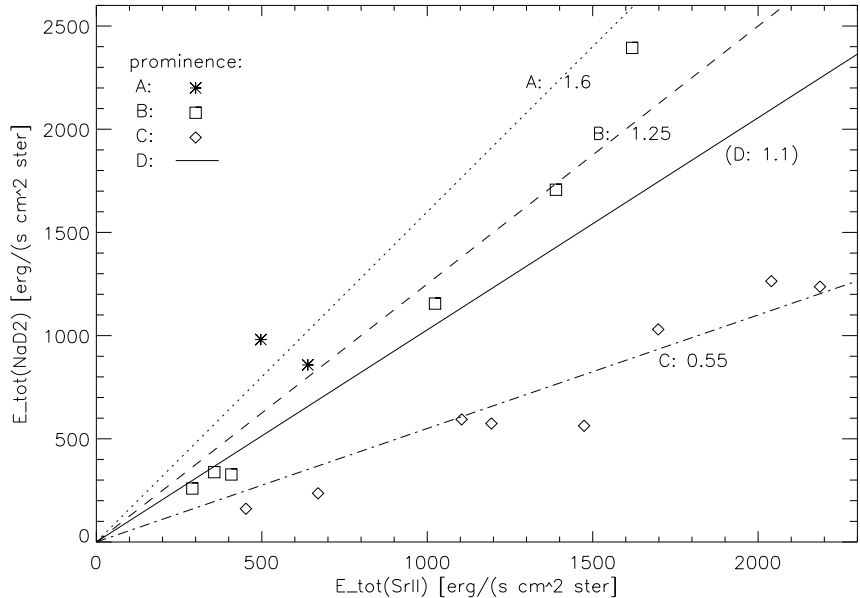} 
   \caption{Integrated line emission, $E_{\rm tot}$, of Na\,D$_2$ and
   Sr\,{\sc ii}\,4078\,\AA{} for prominences A, B and C. Dotted, 
   dashed and dash-dotted lines give the corresponding linear fits; for 
   prominence\,D we plot the mean slope of 1.1 (thick) from Figure\,3.} 
%\label{Fig2}
    \end{figure}
%________________________________________________________________

Spectra showing a good correlation of Na\,D$_2$ and Sr\,{\sc ii} yield 
for the prominences A, B and C the mean line emission ratios 
E(Na\,{\sc i})/E(Sr\,{\sc ii}) of 1.6, 1.25 and 0.55, respectively 
(see Figure\,2). With the relations of Landman (1983\,a) calculated for 
kinetic $T_{\rm kin}=8000$\,K and non-thermal $V_{\rm nth}=3$\,km/s broadening 
parameters, and their correction following Landman (1986) by a factor 
of 0.5, we obtain for these ratios mean electron densities of $n_e$=7, 5 
and 2$\times10^{10}$\,cm$^{-3}$ for prominences A, B and C (Table\,2).
 
Prominence D is observed with an extended coverage of slit-positions and 
on two successive days. This allows us to study the spatial and temporal 
variation of the $E_{\rm tot}$ ratio and thus of $n_e$. Figure 3 shows for 
September 3 an internal scatter of the ratio with a mean of 1.1 corresponding 
to $n_e\approx4.2\times10^{10}$\,cm$^{-3}$. We find no systematic dependence on 
height above the solar limb or on the Balmer brightness $E_{\rm tot}$(H$\delta$). 
On September 2, the mean emission ratio of 1.0 indicates no marked time variation 
of $n_e$ with the prominence evolution over 24 hours.

Although the exact knowledge of $T_{\rm kin}$ and $V_{\rm nth}$ is of no great
importance (Landman 1983a), the systematic ambiguity of these values (see 
Section\,3.2) introduces an uncertainty. Its amount is illustrated by 
comparison with Landman's (1983a) calculations for $T_{\rm kin}=5000$\,K and 
$V_{\rm nth}=5$\,km/s (brackets in Table\,2), which gives about $20\%$ lower 
$n_e$ values than calculations for $T_{\rm kin}=8000$\,K and 
$V_{\rm nth}=3$\,km/s (see Table\,2).

%________________________________________________________________
%  Fig.3  [E_Delta./.E_Sr]
   \begin{figure}
   \hspace{-5mm}
   \includegraphics[width=\textwidth]{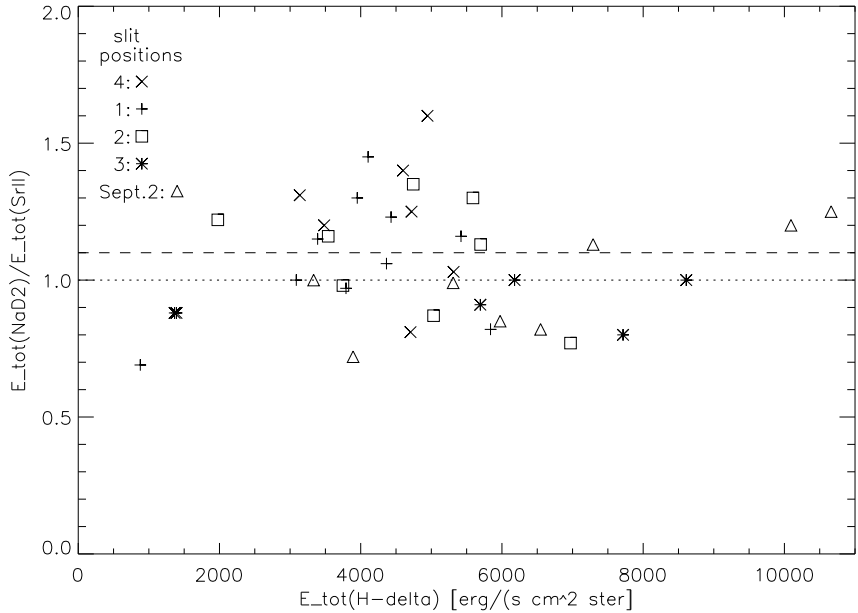} 
   \caption{Ratio of integrated line emissions $E_{\rm tot}$(Na\,D$_2$)
            and $E_{\rm tot}$(Sr\,{\sc ii}\,4078) versus $E_{\rm tot}$(H$\delta$)
            for prominence D on September 3 at 4 slit positions decreasing with 
            height above the limb (see Figure 1), giving a mean of 1.1 (dashed 
            line), and for September 2 with a mean of 1.0 (dotted line).} 
%\label{Fig3}
    \end{figure}
%________________________________________________________________

%+++++++++++++++++++++++++++++++
%% Table(2)
\begin{table}
\caption{Emission ratio of Na\,D$_2$ and Sr\,{\sc ii}\,4078\,\AA{} and
electron density $n_e[10^{10}$cm$^{-3}$] deduced from the diagrams by
Landman (1983a, 1986) for $T_{\rm kin}=8000$\,K and $V_{\rm nth}=3$\,km/s; the 
values for $T_{\rm kin}=5000$\,K and $V_{\rm nth}=5$\,km/s are in parentheses; 
columns\,4 and 5 give the reduced Doppler widths 
$\Delta\lambda_{\rm D}/\lambda_0$\,$[10^{-5}]$ of Na\,D, 
Sr\,{\sc ii} and H$\delta$ for the prominences listed in Table\,1.} 
\begin{tabular}{cccccccc} % l: left, c: center, r: right
\hline 
\hspace{-4mm} prominence&date&E (Na/Sr)&$n_e$&width (Na)&width (Sr)&width (H)\\ 
\hline 
\hspace{-4mm} A&Aug. 30 & 1.6  &7.0 (5.0)  & 1.2  & 1.1  & 3.8  \\
\hspace{-4mm} B&Aug. 31 & 1.25 &5.0 (4.0)  & 1.63 & 1.85 & 4.0  \\
\hspace{-4mm} C&Sept. 1 & 0.55 &2.0 (1.7)  & 1.55 & 1.82 & 4.3  \\ 
\hspace{-4mm} D&Sept. 2 & 1.0  &4.0 (3.2)  & 1.12 & 1.17 & 3.68  \\
\hspace{-4mm} D&Sept. 3 & 1.10 &4.2 (3.5)  & 1.1  & 1.06 & 3.78  \\
%\hline  
\end{tabular}
\end{table}
%+++++++++++++++++++++++++++++++

\subsection{Line Widths}
  \label{S-width}

The deduction of the electron density $n_e$ from the emission ratio depends 
little on the thermal, $T_{\rm kin}$, and non-thermal, $V_{\rm nth}$, contributions 
to line-broadening. These parameters are commonly obtained from inter-comparison 
of Doppler widths, $\Delta\lambda_{\rm D}$, of optically thin lines ({\it i.e.}, 
with Gaussian profiles) emitted by atoms of different atomic mass $\mu$, 
following
 
\begin{equation}
(V_{\rm D})^2=(c \Delta\lambda_{\rm D}/\lambda_0)^2= 2RT_{\rm kin}/\mu+V_{\rm nth}^2,
\end{equation}

\noindent
where $R$ is the gas-constant and $V_{\rm nth}$ is the non-thermal line 
broadening with a Maxwellian velocity distribution. A necessary condition for 
the validity of this equation is that the lines originate in the same volume 
element. Since blending from unresolved macro-velocities introduces additional 
line-broadening ({\it e.g.}, Gun\'ar {\it at al.}, 2008), we determine 
$\Delta\lambda_{\rm D}$ from those spatial emission maxima in the spectra which 
show symmetric line profiles ({\it cf.} end of Section\,2).

%________________________________________________________________
%  Fig.4       [E_tot vs. I_max Delta, Na, Sr+]
   \begin{figure}[h]
   \hspace{-5mm}
   \includegraphics[width=\textwidth]{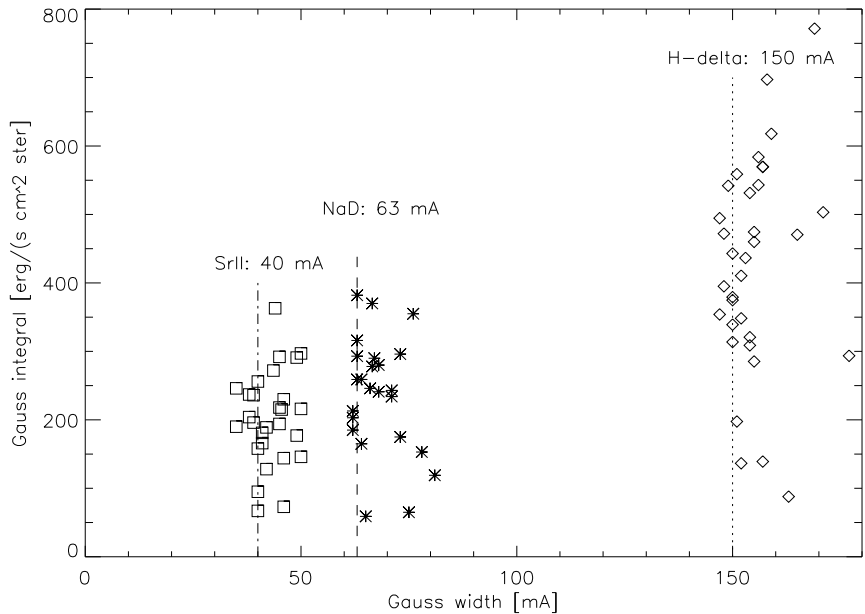}
   \caption{Relation of the width $\Delta\lambda_{\rm e}$ and the 
   integrated intensity $E_{\rm tot}$ of the Gaussian fits of 
   Sr\,{\sc ii}\,4078 (boxes), NaD$_2$ (asterisks) and H$\delta$ (rhombs; 
   scaled to 1/10) for the spatial emission maxima in the four spectra of 
   prominence D shown in Figure 1. The lower means of $\Delta\lambda_{\rm e}$ 
   (vertical lines) are labeled for each emission line with the relevant 
   values, representing the mean $\Delta\lambda_{\rm D}$ over the prominence D.}
%\label{Fig4}
    \end{figure}
%________________________________________________________________

In Figure 4 we plot $\Delta\lambda_{\rm e}$ and $E_{\rm tot}$ from the 
Gaussian fits of the three emission lines in all spectra of the well 
documented prominence D ({\it cf.} Figure 1). We insert a reasonable lower 
mean $\Delta\lambda_{\rm e}^{\rm mean}$ (vertical lines) for each emission 
line, which accounts for additional line broadening (at the right side) and 
statistical noise (at the left side of each mean). We find no significant 
dependence of $\Delta\lambda_{\rm e}$ on the emission brightness $E_{\rm tot}$. 

Assuming Doppler widths $\Delta\lambda_{\rm D}=\Delta\lambda_{\rm e}^{\rm mean}$, 
we obtain for the mostly temperature broadened H$\delta$, 
$\Delta\lambda_{\rm D}=150$\,m\AA{}, which gives, with Equation\,(1), an upper 
limit ({\it i.e.}, $V_{\rm nth}=0$) of $T_{\rm kin}\le7350$\,K. The mostly 
turbulence-broadened Sr\,{\sc ii} line with $\Delta\lambda_{\rm D}=40$\,m\AA{} 
gives an upper limit of the non-thermal broadening ({\it i.e.,} $T_{\rm kin}=0$) 
of $V_{\rm nth}\le3$\,km/s through prominence D. 

Assuming that the neutral lines Na\,D$_2$ and H$\delta$ are formed in the 
same prominence volume, their $\Delta\lambda_{\rm D}$ of 63 and, respectively, 
150\,m\AA{} give T$_{\rm kin}=7050$\,K and $V_{\rm nth}=2.2$\,km/s, well within 
the above limits. Inserting these values into Equation\,(1) we obtain for the 
Sr\,{\sc ii} line a value $\Delta\lambda_{\rm D}=34.3$\,m\AA{}, 
{\it i.e.,} 17\% smaller than the observed $\Delta\lambda_{\rm D}=40$\,m\AA{} 
(Figure 4). Such an excess broadening was already found from our earlier 
observations of Fe\,{\sc ii} and Ti\,{\sc ii} lines (Wiehr {\it et al.}, 2013). 

Similarly, we find that the observed triplet line He\,3888\,\AA{} is broader 
than calculated for $T_{kin}=7050$\,K and $V_{\rm nth}=2.2$\,km/s. This is in 
accordance with earlier findings of a width excess of the triplet line 
He\,4472\,\AA{} (Stellmacher and Wiehr, 2015).

\subsection{Marco-Shifts}
  \label{S-macro}

The excess broadening obtained for lines from ions suggests an interaction 
with the magnetic field ({\it cf.,} Introduction). This seems to be indicated 
from spatial fluctuations of macro-velocities (line-shifts) relative to the 
mean wavelength of each spectrum. For the well documented quiet prominence D 
we find, consistently on both successive days, that the Sr\,{\sc ii} line 
shows 1.27 times larger shifts than the simultaneously observed (neutral) 
H$\delta$ line (Figure\,5). 

%________________________________________________________________
%  Fig.5    [Shift Sr v. Delta]
   \begin{figure}[h]
   \hspace{-5mm}
   \includegraphics[width=\textwidth]{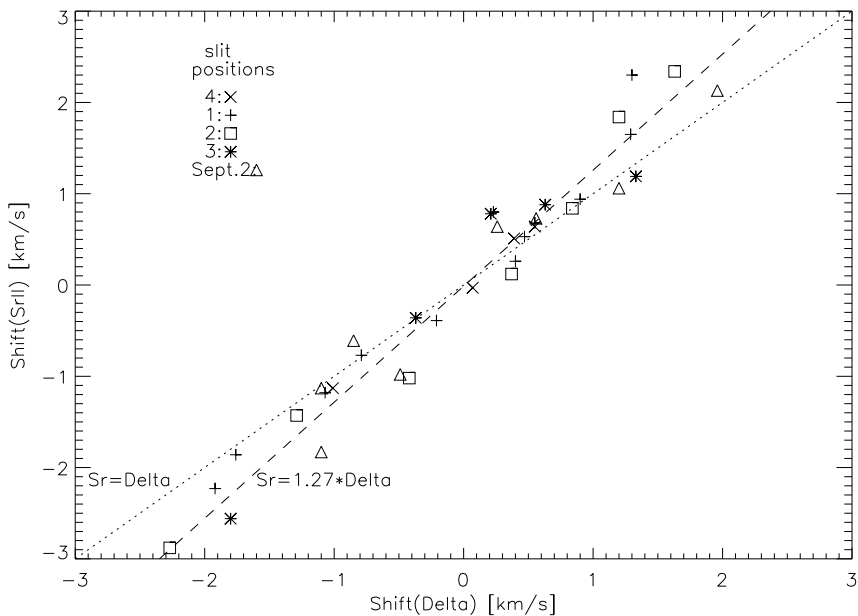}
   \caption{Line shifts of Sr\,{\sc ii}\,4078\,\AA{} versus H$\delta$ relative 
    to the mean wavelengths for prominence D on both days (Table 1). A linear 
    fit gives 1.27 times higher shifts of Sr\,{\sc ii}; the dotted line 
    denotes a 1:1-relation.}
%\label{Fig5}
    \end{figure}
%________________________________________________________________

In order to verify that neutrals show equal macro-shifts (and to get an
idea about the accuracy), we plot in Figure\,6 shifts of the neighboring 
lines He\,{\sc i}\,3888 and H$_8$\,3889\,\AA{}, and find a one-to-one 
relation with small scatter. This is equally indicated for the shifts of 
neutral Na and H$\delta$, although at larger scatter, since these lines 
are not simultaneously observed ({\it cf.,} Section 2). We check whether the 
shift excess of Sr\,{\sc ii} is generally valid for lines from ions using 
the neighboring lines Fe\,{\sc ii}\,5018\,\AA{} and He\,{\sc i}\,5015\,\AA{}, 
earlier observed at the Locarno observatory ({\it cf.,} Stellmacher and Wiehr, 
2015). We also find an excess shift of the (ionized) Fe\,{\sc ii} line over the 
(neutral) He line of the order of 1.3. 

%________________________________________________________________
%  Fig.6    [Shift He vs. H8]
   \begin{figure}
   \hspace{-5mm}
   \includegraphics[width=\textwidth]{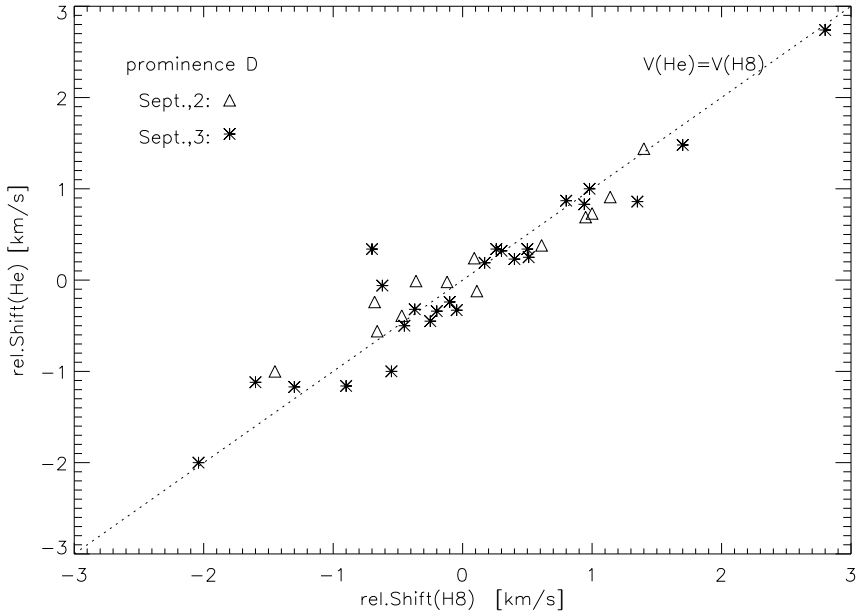}
   \caption{Macro-velocities from He\,{\sc i}\,3888\,\AA{} versus those from
   H$_8$\,3889\,\AA{} relative to their mean wavelengths in prominence\,D 
   on both days.}
%\label{Fig6}
    \end{figure}
%________________________________________________________________

\subsection{Unresolved Prominence Structures}
  \label{S-struct}

The different macro-shifts of neutral and ionized lines indicate
distinct emission regions. These will be of smaller scale than the
spatial resolution achieved, since the spectra (of Na\,{\sc i} and 
Sr\,{\sc ii}) are conspicuously similar (Figure\,1). The macro-shifted 
spectral streaks of $\approx1000$\,km width will result from tight bundles 
of superposing sub-structures. We may estimate the number of elements, 
responsible for the brightness of the Balmer emission within the resolution 
area, adopting for each of them $E$(H$\alpha)=1\times10^4$ erg/(s cm$^2$ ster) 
($\tau_{\alpha}\approx0.1$; {\it cf.,} Stellmacher and Wiehr (2000) and 
references therein). The observed mean H$\delta$ brightness of prominence D, 
$E$(H$\delta)\le8\times10^3$ erg/(s cm$^2$ ster), corresponds in the tables
by Gouttebroze {\it et al.} (1993) to $\tau_{\delta}\le0.09$ and 
$E$(H$\alpha)\le19\times10^4$erg/(s~cm$^2$~ster). The optically thin H$\delta$ 
line will then originate from 19 sub-structures in the resolution area.
These may readily be arranged in a single layer, if smaller than 200\,km, 
in accordance with the narrow streaks in the spectra of prominence\,D, which 
is thus considered to fairly fulfill the conditions for Equation\,(1).  

In contrast, the less structured spectra of the bright prominence C show much 
broader line widths. The nearly six times brighter H$\delta$ emission than 
prominence~D indicates more superposed sub-structures, leading to 
important blending (see Gun\'ar {\it et al.}, 2008), which broadens H$\delta$ 
and H$_8$ to $\Delta\lambda_{\rm D}/\lambda_0=4.3\times10^{-5}$ (as compared to 
prominence\,D with $3.7\times10^{-5}$). Combined with the corresponding width 
of the Na\,D$_2$ line ($\Delta\lambda_{\rm D}/\lambda_0$ = 1.6$\times10^{-5}$), 
Equation\,(1) gives an unrealistically high thermal line broadening of 
$T_{kin}=9080$\,K ($V_{\rm nth}=4$\,km/s) for prominence C. Its total 
thickness of $\tau_{\alpha}\approx25$ and the larger number of superposing 
sub-structures imply that the validity of Equation\,(1) and the deduced 
broadening parameters can hardly be assumed. Besides, bright prominences are 
rather expected to be cooler ({\it cf.,} Stellmacher and Wiehr, 1994a, 1994b, 
1995).
  
\section{Conclusions}
  \label{S-concl}

The similar aspects of the emission maxima of Na\,D$_2$ and 
Sr\,{\sc ii}\,4078\,\AA{} (Figure 1), which have spatial widths of 
2000\,--\,3000\,km, suggest common emission volumes on this scale. 
But the ionized Sr line shows higher non-thermal line broadening and 
macro-shifts than the neutrals (Balmer, Na). This excess is well 
documented for the weaker prominence D, which is observed under good 
seeing conditions. In the brighter prominences (B and C) superposing 
structures along the line-of-sight, will smear the actual macro-shifts.

The finding of excess shifts is hardly compatible with a homogeneous 
(one-component) atmosphere (Terradas {\it et al.}, 2015). Our finding
of a general shift excess of ions suggests that at smaller scales than 
resolved in the spectra, ({\it e.g.} of 150\,km size), distinct volumes 
preferentially emit either lines from ions or from neutrals, as in the 
scenario of Low {\it et al.} (2012). Here, neutral clumps sink through 
the magnetic field until they are stopped when sufficiently re-ionized. 
Vertical motions of neutral clumps, however, cannot explain a line-of-sight 
velocity excess of lines from ions, which may rather be related to the 
stronger coupling to the magnetic field ({\it e.g.} Gilbert, 2011). 
The weak prominence field responds to the ubiquitous motions of its 
photospheric foot-points (Wedemeyer {\it et al.}, 2013; Wedemeyer and 
Steiner, 2014), which will induce macro-velocities. If part of these 
motions is converted into flows or shocks (Hillier {\it et al.}, 2012), 
additional line-shifts and non-thermal line-broadening may occur for 
lines from ions. 

Distinct emission regions, suggested by the macro-shifts, are not in 
accordance with the assumption of a homogeneous prominence atmosphere as 
assumed in the calculations by Landman (1983, 1986) for the estimate of $n_e$ 
from the observed emission ratio of Na\,D$2$ and Sr\,{\sc ii}\,4078\,\AA{}. 
An origin of Na\,D$2$ and Sr\,{\sc ii}\,4078\,\AA{} in different 
small-scale emission regions causes an uncertainty in the deduced
electron density. Since the spectra of both resonance lines are
conspicuously similar at the $\approx1000$\,km scSOLA-D-17-00026.texale (see Figure\,1),
we estimate the influence of their distinct small-scale emission regions 
not to exceed the uncertainty (20\%) of deduced values, based on the different 
parameters ({\it cf.} Table\,2).

The population of the upper Na\,{\sc i}\,D$_2$ level, $^2P_{3/2}$, requires 
the capture of a free electron, which is furnished almost exclusively by 
Hydrogen; $n_e$ then depends on the ionizing UV radiation (mainly Lyman) 
penetrating into the prominence. A possible dependence of the observed 
emission ratio on the Balmer brightness ({\it i.e.,} the type of the prominence, 
{\it cf.} Tables\,1 and 2) is indicated for prominences\,A, B, C, but needs 
further studies. Our observations do not show significant variation of 
the emissions ratio (thus $n_e$) over the prominence\,D, neither with the 
evolution over 24 hours, nor with height above the solar limb.

\begin{acks}
 The authors thank G.\,Monecke for his support with the observations. 
One of us (E.W.) obtained financial support by KIS. 
\end{acks}

\begin{acks} [Disclosure of Potential Conflicts of Interest]
The authors declare that they have no conflicts of interest.
\end{acks}

\end{article} 


\begin{thebibliography}{}
 
\bibitem[\protect\citeauthoryear{{Bommier} \textit{et al.}}{1994}]{Bommier-et-al-1994}
Bommier,~V. Landi-Degl'Innocenti,~E., Leroy,~J.L., Sahal-Brechot,~S.: 1994, \textit{Solar Phys.}\ \textbf{154}, 231  \url{DOI 10.1007/BF00681098}.

\bibitem[\protect\citeauthoryear{{Gilbert}}{2011}]{Gilbert-2011}
Gilbert,~H.: 2011, AIP Conference Proceedings \textbf{1366}, 5  \url{DOI 10.1063/1.3625583} 

\bibitem[\protect\citeauthoryear{{Gouttebroze} \textit{et al.}}{1993}]{Gouttebroze-et-al-2011} 
Gouttebroze,~P., Heinzel,~P., Vial,~J.-C.: 1993, {\aaps} \textbf{99}, 513

\bibitem[\protect\citeauthoryear{{Gunar} \textit{et al.}}{2008}]{Gunar-et-al-2008} 
Gun\'ar,~S., Heinzel,~P., Anzer,~U., Schmieder,~B.: 2008, {\aap} \textbf{490}, 307 \url{DOI 10.1051/0004-6361:200810127}

\bibitem[\protect\citeauthoryear{{Hillier} \textit{et al.}}{2012}]{Hillier-et-al-2012}
Hillier,~A., Berger,~Th., Isobe,~H., Shibata,~K.: 2012, {\apj} \textbf{746}, 120 \url{DOI 10.1088/0004-637X/746/2/120}

\bibitem[\protect\citeauthoryear{{Koutchmy} \textit{et al.}}{1983}]{Koutchmy-et-al-1983} 
Koutchmy,~S., Lebecq,~Ch., Stellmacher,~G.: 1983 {\aap} \textbf{119}, 261

\bibitem[\protect\citeauthoryear{{Labrosse} \textit{et al.}}{2010}]{Labrosse-et-al-2010}  
Labrosse,~P., Heinzel,~P., Vial,~J.-C., Kucera,~T., Parenti,~S., Gun\'ar,~S., Schmieder,~B., Kilper,~G.: 2010, \ssr \textbf{151}, 243 \url{DOI 10.1007/s11214-010-9630-6}

\bibitem[\protect\citeauthoryear{{Labs and Neckels}}{1970}]{Labs-Neckels-1970}
Labs,~D., Neckel,~H.: 1970, {\solphys} \textbf{17}, 50

\bibitem[\protect\citeauthoryear{{Landman} \textit{et al.}}{1978}]{Landman-et-al-1973}  
Landman,~D.A., Illing,~M.E., Mongillo,~M.: 1978, {\apj} \textbf{220}, 666 

\bibitem[\protect\citeauthoryear{{Landman}}{1983}]{Landman-1983a}
Landman,~D.A.: 1983\,a, {\apj} \textbf{269}, 728

\bibitem[\protect\citeauthoryear{{Landman}}{1983}]{Landman-1983b}
Landman,~D.A.: 1983\,b, {\apj} \textbf{270}, 265

\bibitem[\protect\citeauthoryear{{Landman}}{1986}]{Landman-1986}
Landman,~D.A.: 1986, {\apj} \textbf{305}, 546

\bibitem[\protect\citeauthoryear{{Low} \textit{et al.}}{2012}]{Low-et-al-2012} 
Low,~B.C., Liu,~W., Berger,~T., Casini,~R.: 2012, {\apj} \textbf{757}, L21 \url{DOI 10.1088/0004-637X/757/1/21}

\bibitem[\protect\citeauthoryear{{Ramelli} \textit{et al.}}{2012}]{Ramelli-et-al-2012} 
Ramelli,~R., Stellmacher,~G., Wiehr,~E., Bianda,~M.: 2012, {\solphys} \textbf{281}, 697 \url{DOI 10.1007/s11207-012-0118-2}

\bibitem[\protect\citeauthoryear{{Shih-Huei}}{1961}]{Shih-Huei-1961} 
Shih-Huei,~Y.: 1961, Publications of the Crimean Astrophysical Observatory \textbf{25}, 180

\bibitem[\protect\citeauthoryear{{Stellmacher}}{1969}]{Stellmacher-1969}  
Stellmacher,~G.: 1969, {\aap} \textbf{1}, 62 
 
\bibitem[\protect\citeauthoryear{{Stellmacher and Wiehr}}{1994}]{Stellmacher-Wiehr-1994a} 
Stellmacher,~G., Wiehr,~E.: 1994a, {\aap} \textbf{286}, 302 

\bibitem[\protect\citeauthoryear{{Stellmacher and Wiehr}}{1994}]{Stellmacher-Wiehr-1994b}  
Stellmacher,~G., Wiehr,~E.: 1994b, {\aap} \textbf{290}, 655 

\bibitem[\protect\citeauthoryear{{Stellmacher and Wiehr}}{1995}]{Stellmacher-Wiehr-1995} 
Stellmacher,~G., Wiehr,~E.: 1995, {\aap} \textbf{299}, 921 

\bibitem[\protect\citeauthoryear{{Stellmacher and Wiehr}}{2000}]{Stellmacher-Wiehr-2000}  
Stellmacher,~G., Wiehr,~E.: 2000, {\solphys} \textbf{196}, 357 \url{DOI 10.1023/A:1005237823016}

\bibitem[\protect\citeauthoryear{{Stellmacher and Wiehr}}{2015}]{Stellmacher-Wiehr-2015}  
Stellmacher,~G., Wiehr,~E.: 2015, {\aap} \textbf{581}, 141 \url{DOI 10.1051/0004-6361/201322781}

\bibitem[\protect\citeauthoryear{{Terradas} \textit{et al.}}{2015}]{Terradas-et-al-2015}  
Terradas,~J., Soler,~R., Oliver,`R., Ballester,~J.L.:  2015, {\apj} \textbf{802}, L28 \url{DOI 10.1088/2041-8205/802/2/L28}

\bibitem[\protect\citeauthoryear{{Wedemeyer} \textit{et al.}}{2013}]{Wedemeyer-et-al-2013}             
Wedemeyer,~S., Scullion,~E., Rouppe van der Voort,~L., Bosnjak,~A., Antolin,~P.: 2013, {\apj} \textbf{774}, 123 \url{DOI 10.1088/0004-637X/774/2/123}

\bibitem[\protect\citeauthoryear{{Wedemeyer and Steiner}}{2014}]{Wedemeyer-Steiner-2014} 
Wedemeyer,~S., Steiner,~O.: 2014, {\pasj} \textbf{66}, 10 \url{DOI 10.1093/pasj/psu086}

\bibitem[\protect\citeauthoryear{{Wiehr} \textit{et al.}}{2013}]{Wiehr-et-al-2013}                 
Wiehr,~E., Stellmacher,~G., Ramelli,~R, Bianda,~M: 2013, Central European Astrophysical Bulletin \textbf{37}, 487 

\bibitem[\protect\citeauthoryear{{Wiehr} \textit{et al.}}{2015}]{Wiehr-et-al-2015}         
Wiehr,~E., Stellmacher,~G.: 2015, Central European Astrophysical Bulletin \textbf{39}, 35 

\bibitem[\protect\citeauthoryear{{Wiehr} \textit{et al.}}{2016}]{Wiehr-et-al-2016}         
Wiehr,~E., Stellmacher,~G., Bianda,~M: 2016, Central European Astrophysical Bulletin \textbf{40}, 79  

\bibitem[\protect\citeauthoryear{{Yakovkin and Zel'dina}}{1964}]{Yakovkin-Zel'dina-1964}  
Yakovkin,~N.A, Zel'dina,~M.Yu. 1964, Soviet\,Astr. {\aj} \textbf{7}, 643

\end{thebibliography}
\end{document}